\documentstyle[12pt]{article}

\newcommand{\degr}{^0}
\begin{document}
\begin{titlepage}
\begin{flushright}
Zurich University ZU-TH 11/95\\
Pavia University FNT/T-95/20\\
\end{flushright}
\vfill

\begin{center}
{\large\bf MACHOs AND MOLECULAR CLOUDS IN GALACTIC HALOS$^{\dagger}$}
\vskip 0.5cm
F.~De Paolis$^{1,2,\star}$,
G.~Ingrosso$^{1,2,\star}$ ,
Ph.~Jetzer$^{3,\star\star}$
and M.~Roncadelli$^{4}$
\vskip 0.5cm
$^1$
Dipartimento di Fisica, Universit\`a di Lecce, Via Arnesano, CP 193,
73100
Lecce, Italy\\
$^2$
INFN, Sezione di Lecce, Via Arnesano, CP 193, 73100 Lecce, Italy\\
$^3$
Institute of Theoretical Physics, University of
Zurich,Winterthurerstrasse
190, CH-8057 Zurich, Switzerland\\
$^4$
INFN, Sezione di Pavia, Via Bassi 6, I-27100, Pavia

\vfill
\end{center}
\begin{abstract}
Recent observations of microlensing events in the Large Magellanic
Cloud by
the MACHO and EROS collaborations suggest that an important fraction of
the
galactic halo is in the form of Massive Astrophysical Compact Halo
Objects
(MACHOs) with mass
$\sim 0.1 M_{\odot}$.
We outline a scenario in which dark clusters of MACHOs
and molecular clouds form in the halo at galactocentric distances
larger
than $\sim 10-20$ kpc, provided baryons are a major constituent of the
halo.
Possible signatures of the presence of molecular clouds in our galaxy
are discussed. We also discuss how molecular clouds as well as MACHOs
can
be observed directly in the nearby M31 galaxy.
\end{abstract}
\vfill
\begin{flushleft}
$^{\dagger}$ To appear in the proceedings of the International Cosmic
Ray
Conference \\
(Rome, August 1995)\\
$^{\star}$ This work has been partially supported by Italian Space
Agency\\
$^{\star\star}$ Supported by the Swiss National Science Foundation\\
\end{flushleft}
\end{titlepage}

\newpage
\section{Introduction}
A fundamental question in astrophysics concerns the nature of the
dark matter in galactic halos. While various
exotic dark matter candidates have been proposed,
present limits coming from primordial
nucleosynthesis still allow a halo made of
ordinary baryonic matter. A viable candidate are
MACHOs which can be detected via the gravitational lens effect.
Assuming a standard spherical halo
model, Alcock et al. {\cite{kn:Alcock1}} have found that MACHOs
contribute
with a fraction $0.19^{+0.16}_{-0.10}$ to the halo dark matter, whereas
their
average mass turns out to be $\sim 0.08 M_{\odot}$.
Thus, the problem arises how to explain the
nature of the remaining amount of dark matter in galactic halos.
We proposed a scenario {\cite{kn:Depaolis}} for the formation of dark
clusters of MACHOs and molecular clouds in the galactic halo, which
can be summarized as follows.

After its initial collapse, the proto galaxy (PG) is expected to be
shock
heated to its virial temperature $\sim 10^6$ K. Since overdense regions
cool
more rapidly than average (by hydrogen recombination), proto globular
cluster
(PGC) clouds form in
pressure equilibrium with diffuse gas. At this stage, the PGC
cloud temperature is $\sim 10^4$ K, its mass and size are
$\sim 10^6 (R/kpc)^{1/2} M_{\odot}$ and $\sim 10~(R/kpc)^{1/2}$ pc,
respectively.
The subsequent evolution of the PGC
clouds will be different in the inner and outer part of the galaxy,
depending
on the decreasing collision rate and ultraviolet (UV) fluxes
as the galactocentric distance increases.
Below $10^4$ K, the main coolants are $H_2$ molecules and any heavy
element
produced in a first chaotic galactic phase.
In the central region of the galaxy
an Active Galactic Nucleus and/or a first population of
massive stars are expected to exist,
which act as strong sources of UV radiation that dissociates
the $H_2$ molecules present in the inner part of the halo. As a
consequence,
cooling is heavily suppressed and so inner PGC clouds remain for a long
time at temperature $\sim 10^4$ K, resulting in the imprinting
of a characteristic mass $\sim 10^6 M_{\odot}$.
Later on, the cloud temperature suddenly drops below
$10^4$ K and the subsequent evolution leads to the formation of stars
and ultimately to stellar globular clusters.
In the outer regions of the halo
the UV-flux is suppressed,
so that no substantial $H_2$ depletion actually happens.
This fact has three distinct implications:
$(i)$ no imprinting of a characteristic PGC cloud mass shows up,
$(ii)$ the Jeans mass can now be lower than $10^{-1} M_{\odot}$,
$(iii)$  the cooling time is much shorter than the collision time.
PGC clouds subsequently fragment into
smaller clouds that remain optically thin until the minimum value
of the Jeans mass is attained, thus
leading to MACHO formation in dark clusters.
Moreover, because the
conversion efficiency of the constituent gas in MACHOs
could scarcely have been 100\%,
we expect the remaining fraction $f$ of the gas to form
self-gravitating
molecular clouds, since, in the absence of
strong stellar winds, the surviving gas
remains bound in the dark cluster, but not in diffuse form as in this
case
the gas would be observable in the radio band.

\section{Observational Tests}
Let us now address the possible signatures of the above scenario, in
addition
to the single MACHO detection via microlensing.

We proceed to estimate the $\gamma$-ray flux produced in
molecular clouds through the interaction
with high-energy cosmic-ray protons.
Cosmic rays scatter on protons in the molecules producing $\pi^0$'s,
which subsequently decay into $\gamma$'s.
An essential ingredient is the knowledge of the cosmic ray
flux in the halo. Unfortunately, this quantity is experimentally
unknown
and the only available information comes from theoretical estimates.
More precisely, from the mass-loss rate of a
typical galaxy we infer a total cosmic ray flux in the halo
$F \simeq 1.1\times 10^{-4}$ erg cm$^{-2}$ s$^{-1}$.
We also need the energy distribution of the
cosmic rays, for which we assume the same energy dependence
as measured on the Earth. We then scale the overall
density in such a way that the integrated energy flux agrees with the
above
value. Moreover, we assume that the cosmic ray density scales as
$R^{-2}$ for large galactocentric distance $R$.
Accordingly, we obtain \cite{kn:Depaolis}
\begin{equation}
\Phi_{CR}(E, R) \simeq 1.9\times 10^{-3}~\Phi_{CR}^{\oplus}(E)~
\frac{a^2+R_{GC}^2}{a^2+R^2}~, \label{eqno:45}
\end{equation}
where $\Phi_{CR}^{\oplus}(E)$ is the measured primary cosmic ray flux
on
the Earth,
$a\sim 5$ kpc is the halo core radius and $R_{GC}\sim 8.5$ kpc is our
distance from the galactic center.
The source function
$q_{\gamma}(r)$, which gives the photon number density at distance
$r$ from the Earth, is
\begin{equation}
q_{\gamma}(r)=\frac{4\pi}{m_p}\rho_{H_2}(r) \int dE_p~
\Phi_{CR}(E_p,R(r))~
\sigma_{in}(p_{lab}) <n_{\gamma}(E_p)>~. \label{eqno:49}
\end{equation}
Actually, the cosmic ray protons in the halo
which originate from the galactic disk are mainly directed outwards.
This
circumstance implies that the induced photons will predominantly
leave the galaxy.
However, the presence of magnetic fields in the halo might give rise
to a temporary confinement of the cosmic ray protons similarly to what
happens in the disk.
In addition, there could also be sources of
cosmic ray protons located in the halo itself, as for instance
isolated or binary pulsars in globular clusters.
As we are unable to give a quantitative estimate of the
above effects, we take them into account by introducing an
efficiency factor $\epsilon$, which could be rather small. In this way,
the $\gamma$-ray photon flux reaching
the Earth is obtained by multiplying
$q_{\gamma}(r)$ by $\epsilon/4\pi r^2$ and integrating
the resulting quantity over the cloud volume along the line of sight.

The best chance to detect the $\gamma$-rays in question is provided
by observations at high galactic latitude. Therefore we
find
\begin{equation}
\Phi_{\gamma}(90^0) \simeq \epsilon f~ 3.5 \times 10^{-6}~
{\rm \frac{photons}{cm^2~ s~sr}}~. \label{eqno:53}
\end{equation}

The inferred upper bound for $\gamma$-rays in the 0.8 - 6 GeV
range at high galactic latitude is $3 \times 10^{-7}$ photons cm$^{-2}$
s$^{-1}$ sr$^{-1}$ {\cite{kn:Bouquet}}.
Hence, we see from eq. (\ref{eqno:53}) that the presence of
halo molecular clouds does not lead nowadays to
any contradiction with such an upper limit,
provided $\epsilon f < 10^{-1}$.

Molecular clouds can be detected via the anisotropy they would
introduce
in the Cosmic Background Radiation (CBR), even if
the ratio of the temperature excess
of the clouds to the CBR temperature is less than $\sim 10^{-3}$.
Consider molecular clouds in M31.
Because we expect they have
typical rotational speeds of
$50~-~100$ km s$^{-1}$, the Doppler shift effect
will show up as an anisotropy in the CBR. The
corresponding anisotropy is then \cite{kn:dijqr}
\begin{equation}
\frac{\Delta T}{T_r}= \pm \frac{v}{c}~S~f~\tau_{\nu}~, \label{1}
\end{equation}
where $S$ is the spatial filling factor
and $T_r$ is the CBR temperature.
If the clouds are optically thick only at some frequencies, one can use
the average optical depth over the frequency range of the detector
$\bar \tau$.
We estimate the expected CBR anisotropy between two fields
of view (on opposite sides of M31) separated by $\sim 4\degr$
and with angular resolution of $\sim 1\degr$.
Supposing that the halo of M31 consists of
$\sim 10^6$ dark clusters and that all of them lie between 25 kpc and
35 kpc,
we would be able to detect $10^{3}-10^{4}$ dark clusters per degree
square. Scanning an  annulus of $1\degr$ width and internal
angular diameter $4\degr$, centered at M31, in 180 steps of $1\degr$,
we
would find anisotropies of $\sim 2 \times 10^{-5}~f~\bar \tau$ in
$\Delta T/T_r$ (as now $S=1/25$). In conclusion, the theory does not
permit to establish whether the expected anisotropy lies above or below
current detectability ($\sim 10^{-6}$), and so only observations
can resolve this issue.

Let us now turn to the possibility of detecting MACHOs in M31 via their
infrared emission. For simplicity, we assume all MACHOs
have equal mass $\sim 0.08~M_{\odot}$ (which is the upper
mass limit for brown dwarfs) and make up the fraction $(1-f)$ of the
dark matter in M31. In addition, we suppose that all MACHOs have
the same age $t\sim 10^{10}$ yr {\cite{kn:ad}}.
As a consequence, MACHOs emit most of their
radiation at the wavelength $\lambda_{max}\sim 2.6~{\rm \mu m}$.
The infrared surface brightness $I_{\nu}(b)$ of the M31 dark halo
as a function of the projected separation $b$ (impact parameter)
is given by
\begin{equation}
I_{\nu}(b) \sim  5\times 10^{5} \frac{x^3}{e^x-1}
\frac{a^2(1-f)}{D\sqrt{a^2+b^2}} \arctan \sqrt { \frac{L^2-b^2}
{a^2+b^2} }~~
{\rm Jy~sr^{-1}}, \label{6}
\end{equation}
where the M31 dark halo radius is taken to be $L \sim 50$ kpc.
Some numerical values of
$I_{\nu_{max}}(b)$ with $b=20$ and $40$ kpc are
$\sim 1.6 \times 10^{3}~(1-f)~{\rm Jy~sr^{-1}}$ and
$\sim 0.4 \times 10^{3}~(1-f)~{\rm Jy~sr^{-1}}$, respectively.
The planned SIRTF Satellite contains an
array camera with expected sensitivity of $\sim 1.7 \times 10^{3}~
{\rm Jy~sr^{-1}}$
per spatial resolution element in the wavelength range 2-6 $\mu$m.
Therefore, the MACHOs in the halo of M31 can,
hopefully, be detected in the near future.

\end{document}